\documentclass[aps,
		prd,
		reprint,
		twocolumn,
		superscriptaddress,
		shortbibliography,
		nofootinbib,
		floatfix,
		notitlepage
		]{revtex4-1}
\pdfoutput=1

\usepackage{slashed}
\usepackage{mathrsfs}
\usepackage{amsmath}
\usepackage{amssymb}
\usepackage{revsymb}
\usepackage{graphicx,accents}
\usepackage{graphicx,epstopdf}
\usepackage{mathrsfs}
\usepackage{bm}
\usepackage{psfrag}
\usepackage{hyperref}
\hypersetup{colorlinks=true,citecolor=Red,urlcolor=Red}
\usepackage{bbm}
\usepackage{bbold}
\usepackage{tabularx}
\usepackage{graphicx}
\usepackage[normalem]{ulem}
\usepackage[usenames,dvipsnames]{xcolor}

\usepackage[usenames,dvipsnames]{xcolor}
\usepackage{multirow,tabularx}
\usepackage{xcolor,pifont}
\newcommand*\colourcheck[1]{%
  \expandafter\newcommand\csname #1check\endcsname{\textcolor{#1}{\ding{52}}}%
}

\newcommand{\be}{\begin{equation}}
\newcommand{\ee}{\end{equation}}
\newcommand{\bea}{\begin{eqnarray}}
\newcommand{\eea}{\end{eqnarray}}

\newcommand{\trm}[1]{\textrm{#1}}

\newcommand{\LCperp}{{\scriptscriptstyle \perp}}

\newcommand{\ud}{\mathrm{d}}
\newcommand{\vtheta}{\vartheta}
\newcommand{\vsigma}{\varsigma}

\definecolor{bk1}{RGB}{0,200,100}

\begin{document}
\title{Arbitrarily polarized $\gamma$ photon source from nonlinear Compton scattering}
\author{Yu Xin}
\author{Zu-dong Zhao}
\affiliation{College of Physics and Optoelectronic Engineering, Ocean University of China, Qingdao, Shandong, 266100, China}

\author{Suo Tang}
\email{tangsuo@ouc.edu.cn}
\affiliation{College of Physics and Optoelectronic Engineering, Ocean University of China, Qingdao, Shandong, 266100, China}
\affiliation{Engineering Research Center of Advanced Marine Physical Instruments and Equipment, Ministry of Education, Qingdao, Shandong, 266100, China}
\affiliation{Qingdao Key Laboratory of Optics and Optoelectronics, Qingdao, Shandong, 266100, China}


\begin{abstract}
We investigate the nonlinear Compton photon source for upcoming laser-particle experiments in the collision scenario of high-energy electron beams and relativistic laser pulses.
The stronger laser field could not only improve the scattering probability but also induce broader photon beam divergence.
To maximize the photon flux in a realistic narrow angular acceptance, we show that a balance must be met for the laser intensity to boost the scattering probability and simultaneously confine its beam divergence.
For the currently experiment-concerned $10~\trm{GeV}$ energy regime, the balanced laser intensity locates in the intermediate intensity regime of $\xi \sim O(1)$.
In these regimes of laser intensity and electron energy, the accepted photons within a relatively narrow acceptance are highly polarized in the state closely relating to the polarization of the laser pulse, which actually implies a potential route to finely control the photon polarization via the laser polarization.
\end{abstract}
\maketitle
%
\section{Introduction}
Highly polarized $\gamma$-photon beams have been broadly used in various fields from photonuclear reaction~\cite{HORIKAWA2014109}, atomic physics~\cite{drescher419westerwalbesloh}, to medical physics~\cite{weeks1997compton}, particle physics~\cite{sarri2015generation}, even to search the new physics beyond the standard model~\cite{ptv176}.
For the modern-day laser-particle experiments~\cite{Abramowicz:2021zja,slacref1,E320_2021,Fedotov:2022ely,RMP2022_045001}, generation of multi-GeV highly polarized $\gamma$-photons is crucial for the measurements of real photon-photon scattering~\cite{ptv176,gies18} and nonlinear Breit-wheeler pair creation~\cite{gao2022optimal}.
Currently, the inverse Compton scattering (ICS) photon source is one of the most feasible $\gamma$-photon sources, which has been widely used in experiments for the generation of collimated, quasi-monoenergetic and highly polarized $\gamma$-photons~\cite{HAJIMA201635}.
The associate physical process is often referred to as linear Compton scattering and can be simply obtained via the collision between a beam of high-energy electrons and a laser pulse with relatively weak intensity~\cite{PRABeams034401}.
The ICS photons are energy tunable below the energy-cutoff $4\gamma^{2}_{p}\omega_{l}$, according to the electron energy $\gamma_{p}m$, laser frequency $\omega_{l}$ and collision geometry~\cite{S1793626810000440,PRSTAB18110701}.

One of the critical limitation of the ICS source is the low flux of $\gamma$-photons as the scattering probability (or cross section) is strongly suppressed by the weak laser intensity~\cite{SSRF}.
Recently, laser pulses with relativistic intensity are proposed to collide with laser wakefield accelerated electron beams for the generation of brilliant $\gamma$-photons~\cite{NPAllopticalNLC}.
With higher laser intensities, the scattering probability are significantly improved~\cite{BenPRA2020}.
However, the strong laser field could deflect the momentum of the electron beam and induce broader beam divergence for the scattered photon~\cite{PRD056025,SeiptPRA033402}, which, in contrast, would decrease the photon brilliance or the photon flux within a finite angular acceptance.
The improved scattering probability may also result in multiple emissions of photons, which could slow down the incident electrons due to radiation recoil effect~\cite{PRX011020,PRX031004}, and further increase the photon beam divergence.
Therefore, to improve the photon brilliance, the laser intensity must be large enough to boost the scattering probability, but not too large to confine effectively the divergence of the photon beam.

With relativistic laser intensities, the nonlinear effect would become more important in the scattering process: multiple laser photons could be scattered to be a single high-energy $\gamma$-photon~\cite{bula96}, which is referred to as nonlinear Compton scattering~\cite{Fedotov:2022ely,RMP2022_045001}, and could largely extend the energy cutoff of the scattered photon to the higher energy region~\cite{BenPRA2020}.
Simultaneously, this nonlinearity could also lead to the decrease of the photon polarization~\cite{BenPRA2020} and affect the polarization dependence of the scattering photon on that of the laser pulse, which could be used to precisely control the polarization of the ICS photon.

In this paper, we investigate the highly-polarized $\gamma$-photon source potentially acquired in the upcoming laser-particle experiments, such as LUXE~\cite{Abramowicz:2021zja} and E320~\cite{slacref1,E320_2021}, via the nonlinear Compton scattering in the typical scenario: a beam of high-energy electrons collides nearly head-on with an intense laser pulse.
The similar collision scenario has been employed to test the quantum nature of radiation reaction~\cite{PRX011020,PRX031004}.
We show that for the currently experiment concerned energy regime of $E_{p}\sim O(10)~\trm{GeV}$, there exists a balanced laser intensity in the intermediate intensity regime of $\xi=|e|E_{l}/mc\omega_{l}\sim O(1)$~\footnote{The same definition is usually given to the parameter $a_{0}$ in the community of laser-plasma interaction and atomic physics.}, to maximize the photon yield and polarization within the experimental acceptance angle in the level of $O(10)~\mu\trm{rad}$, where $E_{l}$ is the intensity of laser field and $m$ ($e$) is the mass (charge) of the electron.
We also show the polarization of the accepted photons could be finely controlled by adjusting the polarization of the laser pulse.

The paper is organised as follows.
We begin in Sec.~\ref{sec2} by outlining the full QED calculation for the nonlinear Compton scattering and presenting the general properties of the scattered photons, and then in Sec.~\ref{sec3} manifest the existence of the optimal laser intensity for the generation of high-flux photon beam, and analyze the dependence of the photon polarization on that of the laser pulse in Sec.~\ref{sec4}.
We conclude in Sec.~\ref{sec5}.
In Appendix.~\ref{Append}, the influence of multi-photon emissions on the optimal laser intensity is analyzed.
Throughout the discussion, the natural units $\hbar=c=1$ is used.

\section{Theoretical model}~\label{sec2}
The electron-laser collision could be characterized by the energy parameter $\eta=k\cdot p/m^2$ and the laser's scaled vector potential $a^{\mu}=|e|A^{\mu}$, where $p^{\mu}$ is the electron momentum, $k^{\mu}$ is the laser wavevector. For weakly focussed pulses, the laser field can be simply described as a plane wave~\cite{DiPiazza2015PRA,DiPiazza2016PRL,PRD013010}, depending only on the laser phase $\phi=k\cdot x$.
Here, for simplicity, the laser wavevector is specified as $k^\mu = \omega_{l} (1,0,0,-1)$.
The scattering process can be parameterised by the three components of the scattered photon momentum: \mbox{$s=k\cdot \ell/k\cdot p$}, the fraction of the lightfront momentum taken by the scattered photon from the incoming electron, and \mbox{$\bm{r}= \bm{\ell}_{\LCperp}/sm-\bm{p}_{\LCperp}/m$}, photon momenta in the plane perpendicular to the laser propagating direction, scaling the angular spread \mbox{$\ell_{x,y}/sm = m\eta/\omega_{l}\tan(\psi_{x,y}/2)$} along the incoming electron momentum.

We apply full QED calculations for the scattering process in order to resolve the balanced laser intensity, which (as we will see later) appears in the intermediate intensity regime of $\xi\sim O(1)$ for the currently experiment concerned energy of $E_{p}\sim O(10)~\trm{GeV}$.
The scattering probability and photons' polarization have be well documented in the literature~\cite{seipt2017volkov,tang2020highly,MRE0196125}, and
can be formulated as
\begin{subequations}
\begin{align}
& \frac{\ud^{3}\trm{P}}{\ud s \ud^2\bm{r}}=\frac{\alpha }{(2\pi\eta)^2} \frac{s }{1-s} \left[h_{s}(\left|\bm{F}\right|^{2} - \trm{Re}\{SR^{*}\}) + \left|\bm{w} \right|^{2}\right],\label{Eq_NLC_Prob_Polar_Angle}\\
&\Gamma_{1}(s,\bm{r})= \frac{\alpha /(2\pi\eta)^2 }{\ud^{3}\trm{P}/\ud s \ud^2\bm{r}} \frac{s}{1-s} \left( |w_{x} |^{2} -  |w_{y} |^{2} \right)\,,\\
&\Gamma_{2}(s,\bm{r})= \frac{\alpha /(2\pi\eta)^2 }{\ud^{3}\trm{P}/\ud s \ud^2\bm{r}} \frac{s}{1-s} \left(w_{y}w^{*}_{x} + w_{x}w^{*}_{y} \right)\,,\\
&\Gamma_{3}(s,\bm{r})= \frac{\alpha /(2\pi\eta)^2 }{\ud^{3}\trm{P}/\ud s \ud^2\bm{r}} \frac{s}{1-s} i\left(w_{x}w^{*}_{y} - w_{y}w^{*}_{x} \right)g_{s}\,,
\end{align}
\label{Eq_NLC}
\end{subequations}
$\!\!$calculated in terms of the phase integrals~\cite{ILDERTON2020135410}
\begin{subequations}
\begin{align}
	R&=\int\!\ud\phi \, \bigg[1-\frac{l\cdot \pi_{p}(\phi)}{l\cdot p}\bigg]~e^{i\Phi(\phi)}\,,\\
	\bm{F} &= \frac{1}{m}\int\!\ud \phi ~ \bm{a}(\phi)~e^{i\Phi(\phi)}\,, \\
	S~&= \frac{1}{m^2}\int\!\ud\phi~ \bm{a}^{2}(\phi)~e^{i\Phi(\phi)}\,,
\end{align}
\label{Eq_cal}
\end{subequations}
$\!\!$where $\alpha$ is the fine structure constant, $\trm{Re}\left\{\cdot\right\}$ means the real part of the argument, $h_{s} = s^{2}/[2(1-s)]$, $g_{s} =1 + s^{2}/[2(1-s)]$, and $\bm{w} \equiv (w_{x},w_{y}) = \bm{r} R - \bm{F}$ denotes the polarization contribution in the amplitude level, depending on the photon transverse momentum $\bm{r}$ and field polarization $\bm{F}$. The exponent \mbox{$\Phi(\phi)=\int^{\phi}_{\phi_{i}}\ud\phi'\ell\cdot \pi_{p}(\phi')/[(1-s)\eta m^{2}]$} is the phase integral starting from $\phi_{i}$ at which the pulse turns on, \mbox{$\pi^{\mu}_{p}= p^{\mu}+a^{\mu}- k^{\mu}[p\cdot a/k\cdot p + a^{2}/(2k\cdot p)]$} is the electron's instantaneous momentum in a plane laser field.
The spin of the recoiled (initial) electron has been summed (averaged).

The photon polarization is described with three Stokes parameters:
$\Gamma_{1}$ is the degree of linear polarization denoting the polarization preponderance ($|w_{x} |^{2}$) along the $x$-axis over that ($|w_{y}|^{2}$) along $y$-axis,
$\Gamma_{2}$ is also linear polarization giving the preponderance ($|w_{x}\cos45^{\circ} + w_{y}\sin45^{\circ}|^{2}$) along $45^{\circ}$ to $x$-axis over that ($|w_{x}\cos135^{\circ} + w_{y}\sin135^{\circ}|^{2}$) along $135^{\circ}$ to $x$-axis,
and $\Gamma_{3}$ is the circular polarization degree relating to the field rotation as $(w_{x}, w_{y})\times (w^{*}_{x}, w^{*}_{y})$.
The total polarization degree of the scattered photon is calculated as $\Gamma = (\Gamma^2_{1}+\Gamma^2_{2}+\Gamma^2_{3})^{1/2}$.
The angular (energy) dependence of the photon polarization $\Gamma_{1,2,3}(r_{x}, r_{y})$ [$\Gamma_{1,2,3}(s)$] can also be acquired from~(\ref{Eq_NLC}) by integrating the other variable $s$ ($\bm{r}$) and normalizing with $\ud^{2}\trm{P}/\ud r_{x}\ud r_{y}$ ($\ud \trm{P}/\ud s$).

The laser vector potential has the form
\begin{align}
a^{\mu}(\vtheta,\phi)=m\xi~\trm{Re}\left\{\left[0,a_x(\vtheta),a_y(\vtheta),0\right]e^{-i\phi}\right\}f(\phi)\,,
\label{Vector potential}
\end{align}
where the pulse envelope is $f(\phi)=\cos^2[\phi/(2N)]$ in $\left|\phi\right|<N\pi$ and $f(\phi)=0$ otherwise, $a_x(\vtheta)=\cos \vtheta - i \delta \sin \vtheta$, $a_y(\vtheta)=\sin \vtheta + i\delta \cos \vtheta $.
$|\delta|$ denotes the ellipticity of the laser pulse: $|\delta|=0,~1$ corresponds, respectively, to the linearly and circularly polarized laser background and \mbox{$0<|\delta|<1$} gives a laser pulse with the elliptical polarization, and the sign of $\delta$ characterizes the rotation of laser field: $\delta/|\delta|=1$ means the left-hand rotation and $\delta/|\delta|=-1$ is right-hand rotation.
The semi-major axis of the elliptical laser field is along ($\cos\vtheta,~\sin\vtheta$) with the deflection angle $\vtheta\in[-\pi,~\pi]$ in the transverse plane.
The polarization of the laser field can be described by the classical Stokes parameters $(\varsigma_1,~\varsigma_2,~\varsigma_3)$~\cite{jackson1999classical} as
\begin{align}
\varsigma_1 = \frac{1-\delta^2}{1+\delta^2}\cos2\vtheta,~
\varsigma_2 = \frac{1-\delta^2}{1+\delta^2}\sin2\vtheta,~
\varsigma_3 = \frac{2\delta}{1+\delta^2}
\label{Eq_laser_pola}
\end{align}
satisfying the relation $\varsigma_1^{2} + \varsigma_{2}^{2} + \varsigma^{2}_{3}=1$.

\begin{figure}[t!!!]
\includegraphics[width=0.48\textwidth]{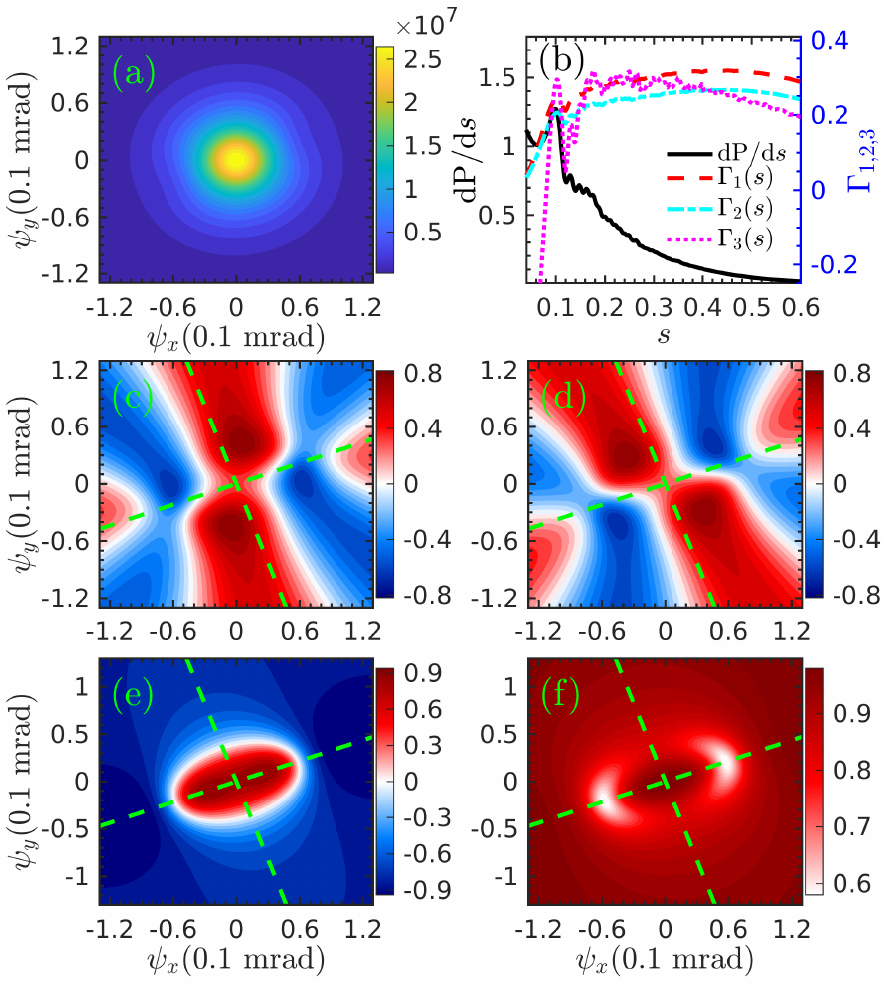}
\caption{The general properties of the $\gamma$-photons scattered by a high-energy ($E_{p}=16.5~\trm{GeV}$) electron colliding with an elliptically polarized laser pulse with ellipticity $\delta=0.5$, deflection angle $\vtheta=\pi/9$, laser duration $N=16$ cycles and intensity $\xi=2$ corresponding to $I=\xi^{2}(1+\delta^{2})/2=2.5$.
The angular distribution $\ud^{2}\trm{P}/\ud\psi_{x}\ud\psi_{y}$ (a) and energy spectrum $\ud\trm{P}/\ud s$ (b) of the scattered photons are given with the angular distributions of the polarizations $\Gamma_{1,2,3}(\psi_{x},\psi_{y})$ in respectively (c)-(e) and the corresponding spectral distributions $\Gamma_{1,2,3}(s)$ in (b) with the blue right $y$-axis.
The total polarization $\Gamma(\psi_{x},\psi_{y})=(\Gamma^2_{1}+\Gamma^2_{2}+\Gamma^2_{3})^{1/2}$ of the scattered photons are given in (f).
The green dashed lines in (c)-(f) denote the directions of the major and minor axis of the field ellipticity.}
\label{Fig1_dist}
\end{figure}

In Fig.~\ref{Fig1_dist}, we plot the general properties of the photons scattered by a high-energy electron with $E_{p}=16.5~\trm{GeV}$
corresponding to the energy parameter $\eta=0.195$ and laser wavelength $\lambda_{l}=0.8~\mu\trm{m}$~\cite{Abramowicz:2021zja}, in the head-on collision with an elliptically polarized laser pulse with $\xi=2$, $\delta=0.5$, $\vtheta=\pi/9$ and $N=16$ cycles.
As shown, in this intermediate intensity region, the photons are scattered in the direction along the electron incident momentum with a finite angular divergence $\sim \xi/\gamma_{p}$~\cite{PRA2020012505,PoP093903} in Fig.~\ref{Fig1_dist}~(a),
and lots of photons are scattered into the high energy region $s>s_{1}=0.1$ with clear harmonic edges in the photon energy spectrum (black solid line) around $s_{n}=2n\eta/(2n\eta+1+I)$ in Fig.~\ref{Fig1_dist}~(b),
where $n$ is the number of the scattered net laser photons~\cite{BenPRA2020},
and $I=\xi^{2}(1+\delta^{2})/2$ is the normalized laser strength, corresponding to the standard intensity of
$2.73\times10^{18} (\lambda_{l}/\mu\trm{m})^{-2} I~\trm{W/cm}^{2}$.
In linearly polarized backgrounds, the angular divergence is about $\sim1/\gamma_{p}$ in the direction perpendicular to the field polarization~\cite{PoP093903}.

The spectral polarization $\Gamma_{1,2,3}(s)$ of the photons depends sensitively on their energy shown in Fig.~\ref{Fig1_dist}~(b) with the blue right-hand vertical axis. In the lower energy region $s<s_{1}$, the polarization degree decreases with $s\to0$ as they are scattered with a larger transverse momentum~\cite{MRE0196125}, and in the high energy region $s>s_{1}$ where the nonlinear effect is dominant, the photons are highly collimated along the electron momentum and possesses considerable polarization relating to the polarization of the laser pulse. 
For small-angle scatterings around $r_{x,y}\approx\psi_{x,y}\to 0$, the polarization contributions $w_{x,y}$ in Eq.~(\ref{Eq_NLC}) comes dominantly from the field polarization as $w_{x,y} \approx F_{x,y}$, as shown in Figs.~\ref{Fig1_dist}~(c)-(f), $\Gamma_{1} = 0.35$, $\Gamma_{2}=0.30$ and $\Gamma_{3}=0.87$ proportional respectively to the laser polarization $\vsigma_{1}=0.46$, $\vsigma_{2}=0.39$ and $\vsigma_{3}=0.80$, which implies the possible control of the photon polarization via the laser polarization.
For the photons with larger scattering angles, their polarization relies not only on the laser polarization, see the green dashed lines denoting the major and minor axis of the field ellipticity in Fig.~\ref{Fig1_dist}~(c)-(d), but also on the transverse momentum of the photons, see Eq.~(\ref{Eq_NLC}).
After integrating over a broad transverse momentum, \emph{i.e.}, for the photon detector (or interaction area) with a broad angular acceptance, the averaged polarization degree would become much smaller than the central polarization as shown in Fig.~\ref{Fig1_dist} (b).

We also note that even though the photon's polarization state $\Gamma_{1,2,3}(\psi_{x},\psi_{y})$ is strongly modulated with different transverse momentum in Fig.~\ref{Fig1_dist}~(c)-(d), the photons with large scattering angles are almost fully polarized with the total degree $\Gamma(\psi_{x},\psi_{y})=(\Gamma^{2}_{1}+\Gamma^{2}_{2}+\Gamma^{2}_{3})^{1/2}>95\%$ in Fig.~\ref{Fig1_dist}~(f).
This means that fully polarized photons in arbitrary polarization state could be acquired via nonlinear Compton scattering by selecting the photons with specified scattering angles $\psi_{x,y}$.

\section{Optimal intensity regime}~\label{sec3}
One of the main applications of highly polarized $\gamma$-photon beams in modern-day laser-particle experiments~\cite{Abramowicz:2021zja,slacref1,E320_2021} is to measure, \emph{e.g.}, vacuum berifringence~\cite{gies18} and Breit-wheeler pair creation~\cite{gao2022optimal} by colliding with a well focussed laser pulse, which could confine the effective interaction area in the scale of the laser waist $w_{0}$.
In realistic experiment setups~\cite{Abramowicz:2021zja}, the focussed laser beam with the waist $w_{0}\sim O(10)~\mu\trm{m}$ is set downstream to the electron momentum for larger photon flux and the distance between the photon generation and laser beam collision is about $d\sim O(1)~\trm{m}$, the effective angular acceptance for the scattered photons would thus be confined in the level of $\Delta \theta=2w_{0}/d\sim O(10)~\mu\trm{rad}$.

\begin{figure}[t!!!]
\includegraphics[width=0.48\textwidth]{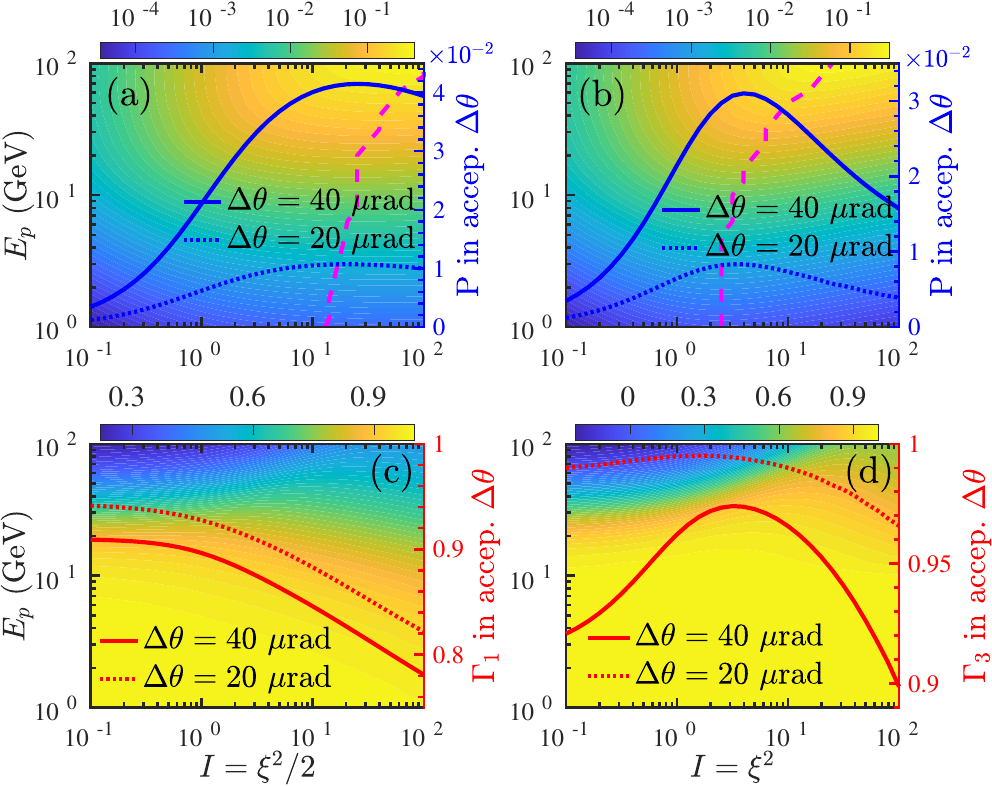}
\caption{Dependence of the photon number $\trm{P}$ [(a) and (b)] and polarization [(c) $\Gamma_{1}$ and (d) $\Gamma_{3}$] on the electron energy $E_{p}$ and laser strength $I$ in linearly [(a) and (c)]  and circularly [(b) and (d)] polarized laser backgrounds for a fixed acceptance angle $\Delta\theta=40~\mu\trm{rad}$.
The magenta dashed lines in (a) and (b) show the laser strength $I_{m}$ for the maximal number of photon in the acceptance from the electron with energy $E_{p}$.
The blue [in (a) and (b) with the right blue $y$-axis] and red [in (c) and (d) with the right red $y$-axis] solid lines indicate the variation of the photon number and polarization with the increase of laser strength $I$ for the fixed electron energy $E_p=16.5~\textrm{GeV}$ and acceptance angle $\Delta\theta=40~\mu\trm{rad}$.
The blue and red dotted lines represent the corresponding results for a smaller acceptance angle $\Delta\theta=20~\mu\trm{rad}$.
The parameters are otherwise the same as Fig.~\ref{Fig1_dist}.}
\label{Fig2_optimal}
\end{figure}

In~Fig.~\ref{Fig2_optimal}, we plot the variation of the photon number $P$ and polarization $\Gamma_{1,3}$ with the change of the electron energy $E_{p}$ and laser strength $I$ for a fixed angular acceptance $\Delta\theta=40~\mu\trm{rad}$, corresponding to $d=1~\trm{m}$ and  $w_{0}=20~\mu\trm{m}$.
As shown in Figs.~\ref{Fig2_optimal} (a) for linearly polarized lasers with $\vsigma_{1}=1$ and (b) circularly polarized lasers with $\vsigma_{3}=1$ , the photon number increases with the increase of the electron energy and laser strength.
For a specified energy $E_{p}$, there exists an optimal strength $I_{m}(E_{p},\Delta \theta)$ which indicates the balance between the boosted scattering probability and beam divergence, see the blue solid lines in Figs.~\ref{Fig2_optimal} (a) and (b) for $E_{p}=16.5~\trm{GeV}$:
In the weak intensity region $I\ll1$, the photon yield improves quickly by increasing the laser strength because of the boosted scattering probability, reaches the maximum in the intermediate laser intensity regime of $\xi \sim O(1)$ at $I_{m}=20$ and $4$ in the linear and circular cases, corresponding respectively to the quantum nonlinear parameter around $\chi=1.2$ and $0.4$, and decreases with the further increased laser strength as the angular divergence of photon beam are largely boosted.
Because the photon beam divergence in linear backgrounds can only be boosted in the direction of the field polarization, the optimal laser intensity in the linear case appears at a larger value than that in the circular case for the same electron energy. 
This difference also leads to the slower decrease of the photon number above the optimal intensity $I_{m}$ in the linearly polarized field in Fig.~\ref{Fig2_optimal} (a) than that in Fig.~\ref{Fig2_optimal} (b) for the circularly polarized field.
For a smaller angular acceptance, the balance intensity would appear at smaller laser strength, see the blue dotted lines in Figs.~\ref{Fig2_optimal} (a) $I_{m}=16$ and $I_{m}=3.2$ in (b) for $\Delta\theta=20~\mu\trm{rad}$.
With the increase of the electron energy $E_{p}$, the optimal intensity $I_{m}$ increases because of the smaller angular spread induced by the field as $r_{x,y}\approx m\eta\psi_{x,y}/2\omega_{l}\approx \xi$ at larger $\eta$, see the magenta dashed lines in Figs.~\ref{Fig2_optimal} (a) and (b).

We can see in Figs.~\ref{Fig2_optimal} (c) and (d) that the detected photons within the narrow acceptance $\Delta\theta=40~\mu\trm{rad}$ are highly polarized, except those at the extreme energy $E_{p}\to 10^{2}~\trm{GeV}$ and low intensity $\xi\to0.1$, as the extreme energy could shift the edge of the first harmonic to the higher energy region $s_{1}\to1$ and the linear scattering with energy $s<s_{1}$ from a single laser photon would dominate the spectrum.
For the electron energy in the currently experiment concerned regime of $E_{p}\sim O(10)~\trm{GeV}$~\cite{Abramowicz:2021zja,slacref1,E320_2021}, see the red solid lines in Figs.~\ref{Fig2_optimal} (c) and (d) for $E_{p}=16.5~\trm{GeV}$ as an example, the photon polarization within the acceptance $\Delta\theta=40~\mu\trm{rad}$ could be higher than $80\%$ in the linear background and $90\%$ in the circular background.
As shown, the polarization degree in the linear background decreases slightly with the increase of the laser strength,
while in the circular background, the polarization degree first increases in the weak intensity region $I<1$ and then decreases quickly with the further increase of laser strength in the region $I>10$.
It is noteworthy that for these $E_{p}\sim O(10)~\trm{GeV}$ electrons, the intensity regime for the relatively large polarization matches well with the optimal intensity region for the photon number.
The photon polarization could be further improved by narrowing the photon acceptance shown as the red dotted lines in Figs.~\ref{Fig2_optimal} (c) and (d) for $\Delta\theta=20~\mu\trm{rad}$.

We point out that for high intensities $I\gtrsim10^{2}$, multiple emission of photons becomes possible when high-energy electrons go through the laser pulse.
The radiation recoil due to photon emissions would reduce the emission of high-energy photons by decelerating the incident electrons,
and increase the beam divergence of the scattered photons because of the finite beaming effect of emissions~\cite{PRA2020012505,TANG2024139136}.
However, for the currently experiment concerned energy regime of $E_{p}\sim O(10)~\trm{GeV}$~\cite{Abramowicz:2021zja,slacref1,E320_2021},
multiple emissions could become considerable for laser strength $I\gtrsim10^{2}$,
and therefore cannot change the optimal strength $I_{m}$ in the intermediate intensity regime, but speed up the decreasing of photon number within the finite acceptance with further increased laser intensity, see discussions in Appendix~\ref{Append}.

\section{Polarization control}\label{sec4}

We can infer from Figs.~\ref{Fig1_dist} (c)-(e) (see the green dashed lines) that the photons' polarization state relates directly to the polarization of the laser pulse, especially for scattering within small angular spread along the electron momentum. 
To manifest the polarization transfer from the laser pulse to the scattered photons, we resort to a monochromatic field $f(\phi)=1$ with the ellipticity $\delta$ and deflection angle $\vtheta$. The exponent $\Phi(\phi)$ in~(\ref{Eq_cal}) can thus be simplified as
\begin{align}
\Phi(\phi)= \kappa \phi + \beta \sin2\phi
\end{align}
with the consideration of central scatterings in a narrow angular spread along the incoming electron momentum, \emph{i.e.}, $r \to 0$,
where $\kappa=s(1 + r^2 + I)/[2\eta (1-s)]$, $\beta = \varsigma_{l} I s/[4\eta (1-s)]$, and $\varsigma_{l} = (1-\delta^2)/(1+\delta^2)$ denotes the total degree of linear polarization.
After doing the Jacobi-Anger expansion for the phase integrands in~(\ref{Eq_cal}) as
\begin{align}
          e^{i\beta\sin(2\phi)} &= \sum_{n=-\infty}^{+\infty} B_{n}(\beta) e^{-in\phi}  \nonumber\\
\cos2\phi~e^{i\beta\sin(2\phi)} &= \sum_{n=-\infty}^{+\infty} D_{n}(\beta) e^{-in\phi}  \nonumber\\
(\cos \phi, \sin \phi)~e^{i\beta\sin(2\phi)} &= \sum_{n=-\infty}^{+\infty} (C_{n}(\beta), iS_{n}(\beta)) e^{-in\phi}  \nonumber
\end{align}
where $B_{n} =(-1)^{n/2}J_{n/2}(\beta)$, $C_{n}=S_{n}=0$, and \mbox{$D_{n} = [J_{-n/2+1}(\beta) + J_{-n/2-1}(\beta)]/2$} if $n$ is a even integer, and if $n$ is an odd integer, {$B_{n} = D_{n} =0$}, \mbox{$C_{n}=[J_{-(n+1)/2}(\beta) + J_{-(n-1)/2}(\beta)]/2$} and \mbox{$S_{n} =[J_{-(n-1)/2}(\beta) - J_{-(n+1)/2}(\beta)]/2$}, and $J_{n}(\beta)$ is the Bessel function of integer order, Eq.~(\ref{Eq_NLC}) can then be rewritten as
\begin{subequations}
\begin{align}
&\frac{\ud^{2}\trm{P}}{\ud\bm{r}^{2}} =\alpha \delta(0)\sum_{n=1}^{+\infty} \frac{4n}{(2n\eta + r^{2}_{*})^{2}}\nonumber\\
&~\left[\left(\xi^{2}C^{2}_{n} + \delta^{2} \xi^{2} S_{n}^{2} - I B^{2}_{n} - I \varsigma_{l} D_{n}B_{n} \right)g_{sn} - B^{2}_{n} \right]\,,\\
&\Gamma_{1}(r_{x},r_{y})=\frac{\alpha \delta(0)}{\ud^{2}\trm{P}/\ud\bm{r}^{2}}\frac{I\vsigma_{1}}{1-\delta^2} \sum_{n=1}^{+\infty} \frac{8n(C^{2}_{n} - \delta^{2} S^{2}_{n})}{(2n\eta + r^{2}_{*} )^{2}}\,,\\
&\Gamma_{2}(r_{x},r_{y})=\frac{\alpha \delta(0)}{\ud^{2}\trm{P}/\ud\bm{r}^{2}}\frac{I\vsigma_{2}}{1-\delta^2} \sum_{n=1}^{+\infty} \frac{8n(C^{2}_{n} - \delta^{2}  S^{2}_{n})}{(2n\eta + r^{2}_{*})^{2}}\,,\\
&\Gamma_{3}(r_{x},r_{y})=\frac{\alpha \delta(0)}{\ud^{2}\trm{P}/\ud\bm{r}^{2}} I\vsigma_{3} \sum_{n=1}^{+\infty}  \frac{8n C_{n}S_{n}}{(2n\eta + r^{2}_{*})^{2}} g_{sr} \,,\label{Eq_NCL_LMA_Ell_G3}
\end{align}
\label{Eq_NCL_LMA_Ell}
\end{subequations}
$\!\!$where $g_{sr} = 1 + s^{2}_{nr}/[2(1-s_{nr})]$, $s_{nr}=2n\eta/(2n\eta+r^{2}_{*})$ and $r^{2}_{*} = \bm{r}^2 + 1 + I$. 
From~(\ref{Eq_NCL_LMA_Ell}), we can clearly see the direct polarization transfer from the laser to the scattered photons $\vsigma_{i}\to \Gamma_{i}$ with ratios $\Gamma_{i}/\vsigma_{i}$ depending on the laser strength $I$ and ellipticity $\delta$.
For a finite laser pulse with a slowly-varying pulse envelope as $f'(\phi)\ll1$, the locally monochromatic approximation for the central scatterings can be acquired via the replacement $\delta(0)\to \int \ud \phi/(2\pi)$ and $\xi \to \xi(\phi)=\xi f(\phi)$~\cite{PoP093903,LMA063110}.
(We have reproduced exactly (not shown here) the central scattered photons with $\psi_{x,y}\to 0$ in Fig.~\ref{Fig1_dist} (a) and (c)-(d) using this locally monochromatic approximation.)

\begin{figure}[t]		
\includegraphics[width=0.48\textwidth]{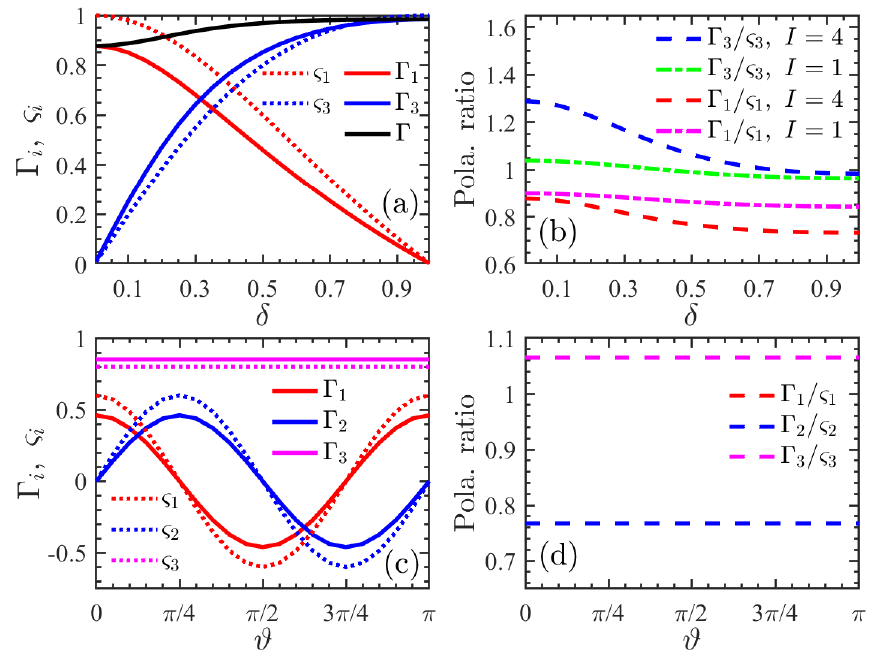}
\caption{Polarization transfer from the laser pulse to the scattered photon beam within a finite angular acceptance $\Delta\theta=40~\mu\trm{rad}$.
(a) The variation of photon polarization with the change of the laser ellipticity from $\delta=0$ to $\delta=1$ for the fixed laser strength $I=4$ and field deflection angle $\vtheta=0$. The total polarization $\Gamma$ (black solid line) of the scattered photon is also attached.
(b) The corresponding transfer ratios $\Gamma_{1}/\vsigma_{1}$ and $\Gamma_{3}/\vsigma_{3}$ are given in comparison with the results from the weaker laser strength $I=1$.
(c) The variation of photon polarization with the change of the field deflection angle from $\vtheta=0$ to $\vtheta=\pi$ with the fixed ellipticity $\delta=0.5$ and strength $I=4$, and the corresponding transfer ratios $\Gamma_{i}/\vsigma_{i}$ are given in (d).
The polarization ratio $\Gamma_{1}/\vsigma_{1}$ (red dashed line) matches exactly with the ratio $\Gamma_{2}/\vsigma_{2}$ (blue dashed line) in (d).
The other parameters are the same as those in Fig.~\ref{Fig1_dist}.}
\label{Fig3_polar}
\end{figure}

We plot in Fig.~\ref{Fig3_polar} the variation of the photon polarization $\Gamma_{1,2,3}$ with the change of the laser polarization $\vsigma_{1,2,3}$ for the electron energy $E_{p}=16.5~\trm{GeV}$ and laser strength $I=4$ with a fixed angular acceptance $\Delta\theta=40~\mu\trm{rad}$.
As shown in Fig.~\ref{Fig3_polar} (a), the polarization degree of the accepted photons varies with the change of the laser ellipticity:
From $\delta=0$ for a linearly polarized laser pulse with $\varsigma_{1}=1$ to $\delta=1$ for a circularly polarized pulse with $\varsigma_{3}=1$, the degree of the photon's linear polarization decreases monotonically from $\Gamma_{1}=0.9$ to $\Gamma_{1}=0$, and its circular polarization degree increases from $\Gamma_{3}=0$ to $\Gamma_{3}=0.97$ monotonically.
The other linear polarization of the photon $\Gamma_{2}$ keeps to be zero because of the laser polarization component $\vsigma_{2}=0$ for $\vtheta=0$.
We can also note in Fig.~\ref{Fig3_polar} (a) that the accepted photons within $\Delta\theta=40~\mu\trm{rad}$ are always highly polarized with the total polarization $\Gamma\gtrsim0.9$ in the variation of laser ellipticity.
For the laser pulse with the ellipticity $\delta\in[-1,0]$, the photon polarization would be consistent with the results for the ellipticity $|\delta|$, except that the photon circular polarization $\Gamma_{3}$ changes its sign since the laser rotation becomes opposite.

In Fig.~\ref{Fig3_polar} (b), we plot the dependence of the polarization ratios $\Gamma_{1}/\vsigma_{1}$ and $\Gamma_{3}/\vsigma_{3}$ between the scattered photons and laser pulse on the laser ellipticity: increasing the laser ellipticity from $\delta=0$ to $\delta=1$ could decrease the linear polarization ratio from $\Gamma_{1}/\vsigma_{1}=0.87$ to $\Gamma_{1}/\vsigma_{1}=0.73$, and from $1.3$ to $0.98$ for the circular polarization ratio $\Gamma_{3}/\vsigma_{3}$.
We note that the circular polarization ratio could be larger than $1$, especially for the small ellipticity $\delta<0.5$.
This is due to the nonlinear effect of multi-photon scattering induced by the strong laser pulse as we can see from~(\ref{Eq_NCL_LMA_Ell}). 
For a weaker laser strength, as also shown in Fig.~\ref{Fig3_polar} (b) for $I=1$, the circular polarization ratio would decrease to around $\Gamma_{3}/\vsigma_{3} = 1$, and simultaneously the linear polarization ratio increases to be slightly larger than $0.85$.
(For ICS photons, the efficiency of the polarization transfer $\Gamma_{3}/\vsigma_{3}$ from the incoming photon to the scattered photon is smaller than $1$~\cite{MRE0196125}.)

Fig.~\ref{Fig3_polar} (c) presents the variation of the photon polarization with the change of field deflection angle $\vtheta$. The two linear polarization vary in the trend as $(\Gamma_{1},~\Gamma_{2})\propto (\cos2\vartheta,~\sin2\vartheta)$, while the circular polarization $\Gamma_{3}$ is constant for different $\vtheta$.
All are in the same trend as the variation of the corresponding laser Stokes parameters in~(\ref{Eq_laser_pola}), which leads to the unchanged polarization ratios $\Gamma_{i}/\vsigma_{i}$ with the change of deflection angle $\vtheta$ as shown in Fig.~\ref{Fig3_polar} (d).
We can thus know that, for the fixed laser ellipticity $\delta$ and intensity $\xi$, the field deflection angle $\vtheta$ can only alter the relative values of the linear polarization with the constant amplitude $(\Gamma^{2}_{1}+\Gamma^{2}_{2})^{1/2}$, but not change the photon's circular polarization $\Gamma_{3}$ and its total polarization $\Gamma$.
For the field deflection angle $\vtheta\in[-\pi,0]$, the polarization $\Gamma_{2}$ changes sign, while the results for $\Gamma_{1}$ and $\Gamma_{3}$ would be the same as those for the deflection $|\vtheta|$.



We point out that the value of the polarization ratio actually determines the sensitivity of the photon polarization depending on the laser polarization.
The large polarization ratios in Fig.~\ref{Fig3_polar} (b) and (d) imply that the polarization of the accepted photons could be finely controlled by adjusting the polarization of the laser pulse, especially for the photon circular polarization as $\Gamma_{3}/\vsigma_{3}\gtrsim 1$.
For the scattered photons within a broader acceptance $\Delta\theta\to \pi$, the polarization ratios would become much smaller as $\Gamma_{i}/\vsigma_{i}<0.5$, see the spectral polarization in Fig.~\ref{Fig1_dist} (b), the polarization control of the photon beam would thus be less sensitive to the laser polarization.

\section{conclusion and discussion}~\label{sec5}

We investigate the nonlinear Compton photon source for the upcoming laser-particle experiments in the typical collision scenario of high-energy electron beams with relativistic laser pulses.
The high-energy $\gamma$-photons are scattered dominantly along the direction of electron momentum within a finite angular divergence $\xi/\gamma_{p}$, proportional to the laser intensity $\xi$ and inversely to the electron energy $\gamma_{p}m$.

With the objective of maximizing the photon flux in realistic experiment setups with the angular acceptance of $\Delta \theta \sim O(10)~\mu\trm{rad}$, a balance must be met for the laser intensity to boost the scattering probability and simultaneously confine the scattered photons' beam divergence.
We show that for the currently experiment-concerned energy regime of $E_{p}\sim O(10)~\trm{GeV}$, the balanced laser intensity locates in the intermediate intensity regime of $\xi \sim O(1)$,
and in these regimes of laser intensity and electron energy, the accepted photons within the narrow acceptance are highly polarized in the polarization state closely relating to the polarization of the laser pulse.
Because of the large polarization ratios $\Gamma_{i}/\vsigma_{i}\approx 1$, the polarization of the accepted photons could be finely controlled by adjusting the laser polarization.

We point out that our conclusion is given for a relatively long laser pulse, and for an ultrashort laser pulse, the photon polarization dependence on the laser Stokes parameters may be different since the laser polarization property could deviate largely from the description of the classical Stokes parameters.
Our conclusions are performed for the scenario of head-on collisions between the electron and laser pulse, while experiments may not allow for perfect head-on collision and there is a small angle between the laser pulse and electron momentum.
This angle will result in a slight reduction to the collision energy parameter $\eta$.

In our consideration, we ignore the angular divergence of the incoming electron beam, which would modify our results considerably if it is much broader than the photon beam divergence induced by the laser field.
The large-angle photons scattered by the high-energy electrons off the beam axis, would decrease the average polarization of the accepted photons downstream the electron beam.
To acquire highly polarized $\gamma$-photon sources, high-quality electron beams with the angular divergence in the same level of or narrower than the laser-induced photon beam divergence are needed.
The other factor that may shift the balanced intensity is the transverse profile of the electron beam and laser pulse as those misaligned electrons would experience lower fields than the laser central field.
However, for the currently experiment-concerned energy regime of $10~\trm{GeV}$, this effect cannot change its optimal intensity significantly, as the laser in the intermediate intensity regime is only weakly focused and the electrons colliding near the pulse center should experience similar fields and contribute dominantly to the scatterings.

\section{Acknowledgments}
Yu Xin and Zu-dong Zhao contribute equally to this work.
The authors acknowledge support from the Shandong Provincial Natural Science Foundation, Grants No. ZR2021QA088.
The work was carried out at Marine Big Data Center of Institute for Advanced Ocean Study of Ocean University of China.

\appendix
\section{Multiple Photon Emission}~\label{Append}
With the increase of the laser intensity $I\gtrsim10^{2}$, the total scattering probability could be boosted to be larger than $1$, and multiple photons could thus be emitted when a high-energy electron goes through the laser pulse.
The radiation recoil from these multiple emissions, could play an important role in the electron's dynamics and affect properties of the scattered photon beam.
The full QED calculations describe the emission process without considering the radiation recoil effect.
To analyze the influence of multi-photon emissions on our conclusion,
we perform single-particle simulations using the standard Monte-Carlo algorithm to simulate the multiple emission process.
The details of our code have been given in Ref.~\cite{TANG2024139136} based on the energy- and angularly resolved locally constant field approximation (LCFA)~\cite{ritus85,2015PRE023305,Piazza2018PRA012134}.

\begin{figure}[t!!!!]		
\includegraphics[width=0.48\textwidth]{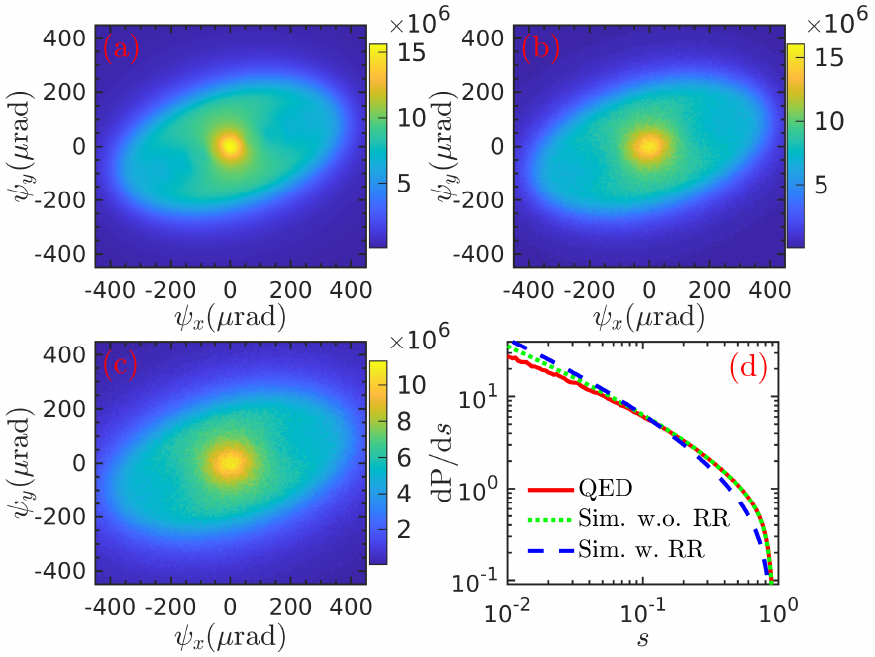}
\caption{Comparison between the full QED result (a) and simulation that does not (b) and does (c) include the radiation recoil (RR) effect in the photon emission.
(a)-(c) Angular distributions of the emitted photons by a high-energy electron, $E_{p}=16.5~\trm{GeV}$, in the head-on collision with an elliptically polarized laser pulse.
(d) The energy spectra of the emitted photon are compared between the QED result and simulation with and without the radiation recoil.
The laser pulse has strength $I=\xi^{2}(1+\delta^{2})/2=100$, ellipticity $\delta=0.5$, field deflection $\vtheta=\pi/9$ and $N=16$ cycles.}
\label{Fig_App1}
\end{figure}

The angular and energy distributions of the scattered photons with and without radiation recoil in the electron's dynamics are compared in Fig.~\ref{Fig_App1}.
The simulation results are averaged over $10^{7}$ electrons with the same incident conditions, and benchmarked against the full QED calculations [Eq.~(\ref{Eq_NLC_Prob_Polar_Angle})]. In simulations, the pair creation by emitted photons is ignored.
As shown, the full QED calculation in Fig.~\ref{Fig_App1} (a) for the photons' angular distribution and energy spectrum (red solid line) in Fig.~\ref{Fig_App1} (d) can be well reproduced by simulations when the radiation recoil is excluded in the electron's dynamics, see the angular distribution in Fig.~\ref{Fig_App1} (b) and photon spectrum (green dotted line) in Fig.~\ref{Fig_App1} (d).
The difference in the low-energy spectra $s\to0$ is because of the infrared divergence of the LCFA~\cite{king19a,PiazzaPRA2019}.

As we can also see, including the radiation recoil would affect considerably the angular and energy distribution of the scattered photon beam.
First, the radiation recoil effect increases the beam divergence of scattered photons as shown in Fig.~\ref{Fig_App1} (c), due to the deceleration of the high-energy electron and the finite beaming of the emissions~\cite{PRA2020012505,TANG2024139136}.
This effect leads to the significant decrease of the photon number along the direction of the electron momentum with $\psi_{x,y}\to 0$.
Second, the radiation recoil reduces (rises) the emission of high (low) energy photons since high-energy electrons can hardly go into the central region of the laser pulse before emitting lots of low-energy photons as shown in Fig.~\ref{Fig_App1} (d).

\begin{figure}[t!!!]	
\includegraphics[width=0.4\textwidth]{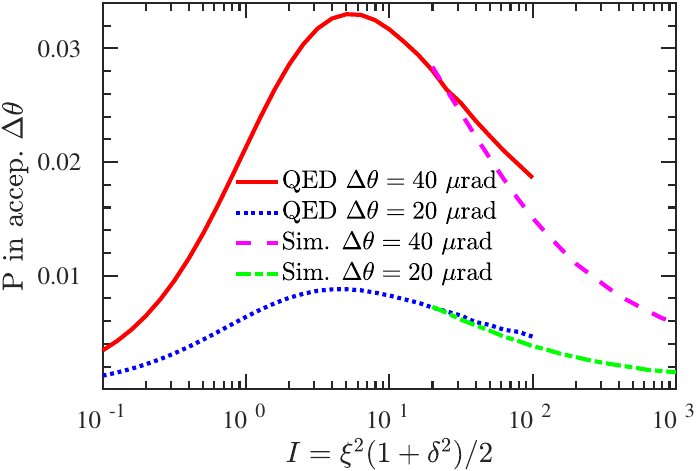}
\caption{Photon yield variation $\trm{P}$ within different angular acceptance $\Delta\theta=20, 40~\mu\trm{rad}$ with the increase of laser strength $I=\xi^{2}(1+\delta^{2})/2$, acquired with the full QED calculations (red solid and blue dotted lines) and simulations including the radiation recoil effect (magenta dashed and green dot-dashed lines).
The other parameters are same as those in Fig.~\ref{Fig_App1}}
\label{Fig_App2}
\end{figure}

In Fig.~\ref{Fig_App2}, we plot the variation of the photon number $P$ with the change of laser strength $I$ for different angular acceptance $\Delta\theta=20~\mu\trm{rad}$, and $40~\mu\trm{rad}$.
Similar as the discussion for Figs.~\ref{Fig2_optimal} (a) and (b), the photon number improves quickly by increasing the laser strength from the weak intensity region $I\ll1$, reaches the maximum at $I_{m}=6.3$ for $\Delta\theta=40~\mu\trm{rad}$ and $I_{m}=5.0$ for $\Delta\theta=20~\mu\trm{rad}$
in the intermediate laser intensity regime, and then decreases with the further increased laser strength.
As we can also see, the effect of radiation recoil starts to be considerable at the strength around $I=40>>I_m$ for the electron energy $E_{p}=16.5~\trm{GeV}$ by speeding up the decreasing of the photon yield, and thus cannot shift the location of the intensities for maximal photon yield within the finite acceptance.

\bibliographystyle{apsrev}

\begin{thebibliography}{41}
\expandafter\ifx\csname natexlab\endcsname\relax\def\natexlab#1{#1}\fi
\expandafter\ifx\csname bibnamefont\endcsname\relax
  \def\bibnamefont#1{#1}\fi
\expandafter\ifx\csname bibfnamefont\endcsname\relax
  \def\bibfnamefont#1{#1}\fi
\expandafter\ifx\csname citenamefont\endcsname\relax
  \def\citenamefont#1{#1}\fi
\expandafter\ifx\csname url\endcsname\relax
  \def\url#1{\texttt{#1}}\fi
\expandafter\ifx\csname urlprefix\endcsname\relax\def\urlprefix{URL }\fi
\providecommand{\bibinfo}[2]{#2}
\providecommand{\eprint}[2][]{\url{#2}}

\bibitem[{\citenamefont{Horikawa et~al.}(2014)\citenamefont{Horikawa, Miyamoto,
  Mochizuki, Amano, Li, Imasaki, Izawa, Ogata, Chiba, and
  Hayakawa}}]{HORIKAWA2014109}
\bibinfo{author}{\bibfnamefont{K.}~\bibnamefont{Horikawa}},
  \bibinfo{author}{\bibfnamefont{S.}~\bibnamefont{Miyamoto}},
  \bibinfo{author}{\bibfnamefont{T.}~\bibnamefont{Mochizuki}},
  \bibinfo{author}{\bibfnamefont{S.}~\bibnamefont{Amano}},
  \bibinfo{author}{\bibfnamefont{D.}~\bibnamefont{Li}},
  \bibinfo{author}{\bibfnamefont{K.}~\bibnamefont{Imasaki}},
  \bibinfo{author}{\bibfnamefont{Y.}~\bibnamefont{Izawa}},
  \bibinfo{author}{\bibfnamefont{K.}~\bibnamefont{Ogata}},
  \bibinfo{author}{\bibfnamefont{S.}~\bibnamefont{Chiba}}, \bibnamefont{and}
  \bibinfo{author}{\bibfnamefont{T.}~\bibnamefont{Hayakawa}},
  \bibinfo{journal}{Physics Letters B} \textbf{\bibinfo{volume}{737}},
  \bibinfo{pages}{109} (\bibinfo{year}{2014}), ISSN \bibinfo{issn}{0370-2693},
  \urlprefix\url{https://www.sciencedirect.com/science/article/pii/S0370269314005917}.

\bibitem[{\citenamefont{Drescher et~al.}(2002)\citenamefont{Drescher,
  Hentschel, Kienberger, Uiberacker, Yakovlev, Scrinzi, Westerwalbesloh,
  Kleineberg, Heinzmann, and Krausz}}]{drescher419westerwalbesloh}
\bibinfo{author}{\bibfnamefont{M.}~\bibnamefont{Drescher}},
  \bibinfo{author}{\bibfnamefont{M.}~\bibnamefont{Hentschel}},
  \bibinfo{author}{\bibfnamefont{R.}~\bibnamefont{Kienberger}},
  \bibinfo{author}{\bibfnamefont{M.}~\bibnamefont{Uiberacker}},
  \bibinfo{author}{\bibfnamefont{V.}~\bibnamefont{Yakovlev}},
  \bibinfo{author}{\bibfnamefont{A.}~\bibnamefont{Scrinzi}},
  \bibinfo{author}{\bibfnamefont{T.}~\bibnamefont{Westerwalbesloh}},
  \bibinfo{author}{\bibfnamefont{U.}~\bibnamefont{Kleineberg}},
  \bibinfo{author}{\bibfnamefont{U.}~\bibnamefont{Heinzmann}},
  \bibnamefont{and} \bibinfo{author}{\bibfnamefont{F.}~\bibnamefont{Krausz}},
  \bibinfo{journal}{Nature} \textbf{\bibinfo{volume}{419}},
  \bibinfo{pages}{803} (\bibinfo{year}{2002}).

\bibitem[{\citenamefont{Weeks et~al.}(1997)\citenamefont{Weeks, Litvinenko, and
  Madey}}]{weeks1997compton}
\bibinfo{author}{\bibfnamefont{K.~J.} \bibnamefont{Weeks}},
  \bibinfo{author}{\bibfnamefont{V.~N.} \bibnamefont{Litvinenko}},
  \bibnamefont{and} \bibinfo{author}{\bibfnamefont{J.~M.~J.}
  \bibnamefont{Madey}}, \bibinfo{journal}{Medical Physics}
  \textbf{\bibinfo{volume}{24}}, \bibinfo{pages}{417} (\bibinfo{year}{1997}),
  \urlprefix\url{https://aapm.onlinelibrary.wiley.com/doi/abs/10.1118/1.597903}.

\bibitem[{\citenamefont{Sarri et~al.}(2015)\citenamefont{Sarri, Poder, Cole,
  Schumaker, Di~Piazza, Reville, Dzelzainis, Doria, Gizzi, Grittani
  et~al.}}]{sarri2015generation}
\bibinfo{author}{\bibfnamefont{G.}~\bibnamefont{Sarri}},
  \bibinfo{author}{\bibfnamefont{K.}~\bibnamefont{Poder}},
  \bibinfo{author}{\bibfnamefont{J.}~\bibnamefont{Cole}},
  \bibinfo{author}{\bibfnamefont{W.}~\bibnamefont{Schumaker}},
  \bibinfo{author}{\bibfnamefont{A.}~\bibnamefont{Di~Piazza}},
  \bibinfo{author}{\bibfnamefont{B.}~\bibnamefont{Reville}},
  \bibinfo{author}{\bibfnamefont{T.}~\bibnamefont{Dzelzainis}},
  \bibinfo{author}{\bibfnamefont{D.}~\bibnamefont{Doria}},
  \bibinfo{author}{\bibfnamefont{L.}~\bibnamefont{Gizzi}},
  \bibinfo{author}{\bibfnamefont{G.}~\bibnamefont{Grittani}},
  \bibnamefont{et~al.}, \bibinfo{journal}{Nature communications}
  \textbf{\bibinfo{volume}{6}}, \bibinfo{pages}{6747} (\bibinfo{year}{2015}),
  \urlprefix\url{https://doi.org/10.1038/ncomms7747}.

\bibitem[{\citenamefont{Homma et~al.}(2016)\citenamefont{Homma, Matsuura, and
  Nakajima}}]{ptv176}
\bibinfo{author}{\bibfnamefont{K.}~\bibnamefont{Homma}},
  \bibinfo{author}{\bibfnamefont{K.}~\bibnamefont{Matsuura}}, \bibnamefont{and}
  \bibinfo{author}{\bibfnamefont{K.}~\bibnamefont{Nakajima}},
  \bibinfo{journal}{Progress of Theoretical and Experimental Physics}
  \textbf{\bibinfo{volume}{2016}}, \bibinfo{pages}{013C01}
  (\bibinfo{year}{2016}), ISSN \bibinfo{issn}{2050-3911},
  \urlprefix\url{https://doi.org/10.1093/ptep/ptv176}.

\bibitem[{\citenamefont{Abramowicz et~al.}(2021)}]{Abramowicz:2021zja}
\bibinfo{author}{\bibfnamefont{H.}~\bibnamefont{Abramowicz}}
  \bibnamefont{et~al.}, \bibinfo{journal}{Eur. Phys. J. Spec. Top.}
  \textbf{\bibinfo{volume}{230}}, \bibinfo{pages}{2445} (\bibinfo{year}{2021}),
  \urlprefix\url{https://doi.org/10.1140/epjs/s11734-021-00249-z}.

\bibitem[{\citenamefont{Meuren}(2019)}]{slacref1}
\bibinfo{author}{\bibfnamefont{S.}~\bibnamefont{Meuren}}, in
  \emph{\bibinfo{booktitle}{Third Conference on Extremely High Intensity Laser
  Physics (ExHILP)}} (\bibinfo{year}{2019}),
  \urlprefix\url{https://conf.slac.stanford.edu/facet-2-2019/sites/facet-2-2019.conf.slac.stanford.edu/files/basic-page-docs/sfqed_2019.pdf}.

\bibitem[{\citenamefont{Salgado et~al.}(2021)\citenamefont{Salgado, Cavanagh,
  Tamburini, Storey, Beyer, Bucksbaum, Chen, Di~Piazza, Gerstmayr, Harsh
  et~al.}}]{E320_2021}
\bibinfo{author}{\bibfnamefont{F.~C.} \bibnamefont{Salgado}},
  \bibinfo{author}{\bibfnamefont{N.}~\bibnamefont{Cavanagh}},
  \bibinfo{author}{\bibfnamefont{M.}~\bibnamefont{Tamburini}},
  \bibinfo{author}{\bibfnamefont{D.~W.} \bibnamefont{Storey}},
  \bibinfo{author}{\bibfnamefont{R.}~\bibnamefont{Beyer}},
  \bibinfo{author}{\bibfnamefont{P.~H.} \bibnamefont{Bucksbaum}},
  \bibinfo{author}{\bibfnamefont{Z.}~\bibnamefont{Chen}},
  \bibinfo{author}{\bibfnamefont{A.}~\bibnamefont{Di~Piazza}},
  \bibinfo{author}{\bibfnamefont{E.}~\bibnamefont{Gerstmayr}},
  \bibinfo{author}{\bibnamefont{Harsh}}, \bibnamefont{et~al.},
  \bibinfo{journal}{New Journal of Physics} \textbf{\bibinfo{volume}{24}},
  \bibinfo{pages}{015002} (\bibinfo{year}{2021}), ISSN
  \bibinfo{issn}{1367-2630},
  \urlprefix\url{http://dx.doi.org/10.1088/1367-2630/ac4283}.

\bibitem[{\citenamefont{Fedotov et~al.}(2023)\citenamefont{Fedotov, Ilderton,
  Karbstein, King, Seipt, Taya, and Torgrimsson}}]{Fedotov:2022ely}
\bibinfo{author}{\bibfnamefont{A.}~\bibnamefont{Fedotov}},
  \bibinfo{author}{\bibfnamefont{A.}~\bibnamefont{Ilderton}},
  \bibinfo{author}{\bibfnamefont{F.}~\bibnamefont{Karbstein}},
  \bibinfo{author}{\bibfnamefont{B.}~\bibnamefont{King}},
  \bibinfo{author}{\bibfnamefont{D.}~\bibnamefont{Seipt}},
  \bibinfo{author}{\bibfnamefont{H.}~\bibnamefont{Taya}}, \bibnamefont{and}
  \bibinfo{author}{\bibfnamefont{G.}~\bibnamefont{Torgrimsson}},
  \bibinfo{journal}{Physics Reports} \textbf{\bibinfo{volume}{1010}},
  \bibinfo{pages}{1} (\bibinfo{year}{2023}), ISSN \bibinfo{issn}{0370-1573},
  \bibinfo{note}{advances in QED with intense background fields},
  \urlprefix\url{https://www.sciencedirect.com/science/article/pii/S0370157323000352}.

\bibitem[{\citenamefont{Gonoskov et~al.}(2022)\citenamefont{Gonoskov,
  Blackburn, Marklund, and Bulanov}}]{RMP2022_045001}
\bibinfo{author}{\bibfnamefont{A.}~\bibnamefont{Gonoskov}},
  \bibinfo{author}{\bibfnamefont{T.~G.} \bibnamefont{Blackburn}},
  \bibinfo{author}{\bibfnamefont{M.}~\bibnamefont{Marklund}}, \bibnamefont{and}
  \bibinfo{author}{\bibfnamefont{S.~S.} \bibnamefont{Bulanov}},
  \bibinfo{journal}{Rev. Mod. Phys.} \textbf{\bibinfo{volume}{94}},
  \bibinfo{pages}{045001} (\bibinfo{year}{2022}),
  \urlprefix\url{https://link.aps.org/doi/10.1103/RevModPhys.94.045001}.

\bibitem[{\citenamefont{Gies et~al.}(2018)\citenamefont{Gies, Karbstein,
  Kohlf\"urst, and Seegert}}]{gies18}
\bibinfo{author}{\bibfnamefont{H.}~\bibnamefont{Gies}},
  \bibinfo{author}{\bibfnamefont{F.}~\bibnamefont{Karbstein}},
  \bibinfo{author}{\bibfnamefont{C.}~\bibnamefont{Kohlf\"urst}},
  \bibnamefont{and} \bibinfo{author}{\bibfnamefont{N.}~\bibnamefont{Seegert}},
  \bibinfo{journal}{Phys. Rev. D} \textbf{\bibinfo{volume}{97}},
  \bibinfo{pages}{076002} (\bibinfo{year}{2018}),
  \urlprefix\url{https://link.aps.org/doi/10.1103/PhysRevD.97.076002}.

\bibitem[{\citenamefont{Gao and Tang}(2022)}]{gao2022optimal}
\bibinfo{author}{\bibfnamefont{Y.}~\bibnamefont{Gao}} \bibnamefont{and}
  \bibinfo{author}{\bibfnamefont{S.}~\bibnamefont{Tang}},
  \bibinfo{journal}{Phys. Rev. D} \textbf{\bibinfo{volume}{106}},
  \bibinfo{pages}{056003} (\bibinfo{year}{2022}),
  \urlprefix\url{https://link.aps.org/doi/10.1103/PhysRevD.106.056003}.

\bibitem[{\citenamefont{Hajima}(2016)}]{HAJIMA201635}
\bibinfo{author}{\bibfnamefont{R.}~\bibnamefont{Hajima}},
  \bibinfo{journal}{Physics Procedia} \textbf{\bibinfo{volume}{84}},
  \bibinfo{pages}{35} (\bibinfo{year}{2016}), ISSN \bibinfo{issn}{1875-3892},
  \bibinfo{note}{proceedings of the International Conference "Synchrotron and
  Free electron laser Radiation: generation and application" (SFR-2016), July 4
  - 7, 2016, Novosibirsk, Russia},
  \urlprefix\url{https://www.sciencedirect.com/science/article/pii/S1875389216303029}.

\bibitem[{\citenamefont{Krafft et~al.}(2023)\citenamefont{Krafft,
  Terzi\ifmmode~\acute{c}\else \'{c}\fi{}, Johnson, and
  Wilson}}]{PRABeams034401}
\bibinfo{author}{\bibfnamefont{G.~A.} \bibnamefont{Krafft}},
  \bibinfo{author}{\bibfnamefont{B.}~\bibnamefont{Terzi\ifmmode~\acute{c}\else
  \'{c}\fi{}}}, \bibinfo{author}{\bibfnamefont{E.}~\bibnamefont{Johnson}},
  \bibnamefont{and} \bibinfo{author}{\bibfnamefont{G.}~\bibnamefont{Wilson}},
  \bibinfo{journal}{Phys. Rev. Accel. Beams} \textbf{\bibinfo{volume}{26}},
  \bibinfo{pages}{034401} (\bibinfo{year}{2023}),
  \urlprefix\url{https://link.aps.org/doi/10.1103/PhysRevAccelBeams.26.034401}.

\bibitem[{\citenamefont{Krafft and Priebe}(2010)}]{S1793626810000440}
\bibinfo{author}{\bibfnamefont{G.~A.} \bibnamefont{Krafft}} \bibnamefont{and}
  \bibinfo{author}{\bibfnamefont{G.}~\bibnamefont{Priebe}},
  \bibinfo{journal}{Reviews of Accelerator Science and Technology}
  \textbf{\bibinfo{volume}{03}}, \bibinfo{pages}{147} (\bibinfo{year}{2010}),
  \eprint{https://doi.org/10.1142/S1793626810000440},
  \urlprefix\url{https://doi.org/10.1142/S1793626810000440}.

\bibitem[{\citenamefont{Petrillo et~al.}(2015)\citenamefont{Petrillo, Bacci,
  Curatolo, Drebot, Giribono, Maroli, Rossi, Serafini, Tomassini, Vaccarezza
  et~al.}}]{PRSTAB18110701}
\bibinfo{author}{\bibfnamefont{V.}~\bibnamefont{Petrillo}},
  \bibinfo{author}{\bibfnamefont{A.}~\bibnamefont{Bacci}},
  \bibinfo{author}{\bibfnamefont{C.}~\bibnamefont{Curatolo}},
  \bibinfo{author}{\bibfnamefont{I.}~\bibnamefont{Drebot}},
  \bibinfo{author}{\bibfnamefont{A.}~\bibnamefont{Giribono}},
  \bibinfo{author}{\bibfnamefont{C.}~\bibnamefont{Maroli}},
  \bibinfo{author}{\bibfnamefont{A.~R.} \bibnamefont{Rossi}},
  \bibinfo{author}{\bibfnamefont{L.}~\bibnamefont{Serafini}},
  \bibinfo{author}{\bibfnamefont{P.}~\bibnamefont{Tomassini}},
  \bibinfo{author}{\bibfnamefont{C.}~\bibnamefont{Vaccarezza}},
  \bibnamefont{et~al.}, \bibinfo{journal}{Phys. Rev. ST Accel. Beams}
  \textbf{\bibinfo{volume}{18}}, \bibinfo{pages}{110701}
  (\bibinfo{year}{2015}),
  \urlprefix\url{https://link.aps.org/doi/10.1103/PhysRevSTAB.18.110701}.

\bibitem[{\citenamefont{Wang et~al.}(2022)\citenamefont{Wang, Fan, Liu, Xu,
  Shen, Ma, Utsunomiya, Song, Cao, Hao et~al.}}]{SSRF}
\bibinfo{author}{\bibfnamefont{H.-W.} \bibnamefont{Wang}},
  \bibinfo{author}{\bibfnamefont{G.-T.} \bibnamefont{Fan}},
  \bibinfo{author}{\bibfnamefont{L.-X.} \bibnamefont{Liu}},
  \bibinfo{author}{\bibfnamefont{H.-H.} \bibnamefont{Xu}},
  \bibinfo{author}{\bibfnamefont{W.-Q.} \bibnamefont{Shen}},
  \bibinfo{author}{\bibfnamefont{Y.-G.} \bibnamefont{Ma}},
  \bibinfo{author}{\bibfnamefont{H.}~\bibnamefont{Utsunomiya}},
  \bibinfo{author}{\bibfnamefont{L.-L.} \bibnamefont{Song}},
  \bibinfo{author}{\bibfnamefont{X.-G.} \bibnamefont{Cao}},
  \bibinfo{author}{\bibfnamefont{Z.-R.} \bibnamefont{Hao}},
  \bibnamefont{et~al.}, \bibinfo{journal}{Nuclear Science and Techniques}
  \textbf{\bibinfo{volume}{33}} (\bibinfo{year}{2022}),
  \urlprefix\url{https://doi.org/10.1007/s41365-022-01076-0}.

\bibitem[{\citenamefont{Mirzaie et~al.}(2024)\citenamefont{Mirzaie, Hojbota,
  Kim, Pathak, Pak, Kim, Lee, Yoon, Lee, Rhee et~al.}}]{NPAllopticalNLC}
\bibinfo{author}{\bibfnamefont{M.}~\bibnamefont{Mirzaie}},
  \bibinfo{author}{\bibfnamefont{C.}~\bibnamefont{Hojbota}},
  \bibinfo{author}{\bibfnamefont{D.}~\bibnamefont{Kim}},
  \bibinfo{author}{\bibfnamefont{V.}~\bibnamefont{Pathak}},
  \bibinfo{author}{\bibfnamefont{T.}~\bibnamefont{Pak}},
  \bibinfo{author}{\bibfnamefont{C.}~\bibnamefont{Kim}},
  \bibinfo{author}{\bibfnamefont{H.~W.} \bibnamefont{Lee}},
  \bibinfo{author}{\bibfnamefont{J.~W.} \bibnamefont{Yoon}},
  \bibinfo{author}{\bibfnamefont{S.}~\bibnamefont{Lee}},
  \bibinfo{author}{\bibfnamefont{Y.-J.} \bibnamefont{Rhee}},
  \bibnamefont{et~al.}, \bibinfo{journal}{Nature Photonics}
  \textbf{\bibinfo{volume}{18}}, \bibinfo{pages}{1212} (\bibinfo{year}{2024}).

\bibitem[{\citenamefont{King and Tang}(2020)}]{BenPRA2020}
\bibinfo{author}{\bibfnamefont{B.}~\bibnamefont{King}} \bibnamefont{and}
  \bibinfo{author}{\bibfnamefont{S.}~\bibnamefont{Tang}},
  \bibinfo{journal}{Phys. Rev. A} \textbf{\bibinfo{volume}{102}},
  \bibinfo{pages}{022809} (\bibinfo{year}{2020}),
  \urlprefix\url{https://link.aps.org/doi/10.1103/PhysRevA.102.022809}.

\bibitem[{\citenamefont{Dai et~al.}(2023)\citenamefont{Dai, Jiang, Jiang,
  Shaisultanov, and Chen}}]{PRD056025}
\bibinfo{author}{\bibfnamefont{Y.-N.} \bibnamefont{Dai}},
  \bibinfo{author}{\bibfnamefont{J.-J.} \bibnamefont{Jiang}},
  \bibinfo{author}{\bibfnamefont{Y.-H.} \bibnamefont{Jiang}},
  \bibinfo{author}{\bibfnamefont{R.}~\bibnamefont{Shaisultanov}},
  \bibnamefont{and} \bibinfo{author}{\bibfnamefont{Y.-Y.} \bibnamefont{Chen}},
  \bibinfo{journal}{Phys. Rev. D} \textbf{\bibinfo{volume}{108}},
  \bibinfo{pages}{056025} (\bibinfo{year}{2023}),
  \urlprefix\url{https://link.aps.org/doi/10.1103/PhysRevD.108.056025}.

\bibitem[{\citenamefont{Seipt et~al.}(2014)\citenamefont{Seipt, Rykovanov,
  Surzhykov, and Fritzsche}}]{SeiptPRA033402}
\bibinfo{author}{\bibfnamefont{D.}~\bibnamefont{Seipt}},
  \bibinfo{author}{\bibfnamefont{S.}~\bibnamefont{Rykovanov}},
  \bibinfo{author}{\bibfnamefont{A.}~\bibnamefont{Surzhykov}},
  \bibnamefont{and}
  \bibinfo{author}{\bibfnamefont{s.}~\bibnamefont{Fritzsche}},
  \bibinfo{journal}{Physical Review A} \textbf{\bibinfo{volume}{91}}
  (\bibinfo{year}{2014}).

\bibitem[{\citenamefont{Cole et~al.}(2018)\citenamefont{Cole, Behm, Gerstmayr,
  Blackburn, Wood, Baird, Duff, Harvey, Ilderton, Joglekar et~al.}}]{PRX011020}
\bibinfo{author}{\bibfnamefont{J.~M.} \bibnamefont{Cole}},
  \bibinfo{author}{\bibfnamefont{K.~T.} \bibnamefont{Behm}},
  \bibinfo{author}{\bibfnamefont{E.}~\bibnamefont{Gerstmayr}},
  \bibinfo{author}{\bibfnamefont{T.~G.} \bibnamefont{Blackburn}},
  \bibinfo{author}{\bibfnamefont{J.~C.} \bibnamefont{Wood}},
  \bibinfo{author}{\bibfnamefont{C.~D.} \bibnamefont{Baird}},
  \bibinfo{author}{\bibfnamefont{M.~J.} \bibnamefont{Duff}},
  \bibinfo{author}{\bibfnamefont{C.}~\bibnamefont{Harvey}},
  \bibinfo{author}{\bibfnamefont{A.}~\bibnamefont{Ilderton}},
  \bibinfo{author}{\bibfnamefont{A.~S.} \bibnamefont{Joglekar}},
  \bibnamefont{et~al.}, \bibinfo{journal}{Phys. Rev. X}
  \textbf{\bibinfo{volume}{8}}, \bibinfo{pages}{011020} (\bibinfo{year}{2018}),
  \urlprefix\url{https://link.aps.org/doi/10.1103/PhysRevX.8.011020}.

\bibitem[{\citenamefont{Poder et~al.}(2018)\citenamefont{Poder, Tamburini,
  Sarri, Di~Piazza, Kuschel, Baird, Behm, Bohlen, Cole, Corvan
  et~al.}}]{PRX031004}
\bibinfo{author}{\bibfnamefont{K.}~\bibnamefont{Poder}},
  \bibinfo{author}{\bibfnamefont{M.}~\bibnamefont{Tamburini}},
  \bibinfo{author}{\bibfnamefont{G.}~\bibnamefont{Sarri}},
  \bibinfo{author}{\bibfnamefont{A.}~\bibnamefont{Di~Piazza}},
  \bibinfo{author}{\bibfnamefont{S.}~\bibnamefont{Kuschel}},
  \bibinfo{author}{\bibfnamefont{C.~D.} \bibnamefont{Baird}},
  \bibinfo{author}{\bibfnamefont{K.}~\bibnamefont{Behm}},
  \bibinfo{author}{\bibfnamefont{S.}~\bibnamefont{Bohlen}},
  \bibinfo{author}{\bibfnamefont{J.~M.} \bibnamefont{Cole}},
  \bibinfo{author}{\bibfnamefont{D.~J.} \bibnamefont{Corvan}},
  \bibnamefont{et~al.}, \bibinfo{journal}{Phys. Rev. X}
  \textbf{\bibinfo{volume}{8}}, \bibinfo{pages}{031004} (\bibinfo{year}{2018}),
  \urlprefix\url{https://link.aps.org/doi/10.1103/PhysRevX.8.031004}.

\bibitem[{\citenamefont{Bula et~al.}(1996)}]{bula96}
\bibinfo{author}{\bibfnamefont{C.}~\bibnamefont{Bula}} \bibnamefont{et~al.}
  (\bibinfo{collaboration}{E144}), \bibinfo{journal}{Phys. Rev. Lett.}
  \textbf{\bibinfo{volume}{76}}, \bibinfo{pages}{3116} (\bibinfo{year}{1996}).

\bibitem[{\citenamefont{Di~Piazza}(2015)}]{DiPiazza2015PRA}
\bibinfo{author}{\bibfnamefont{A.}~\bibnamefont{Di~Piazza}},
  \bibinfo{journal}{Phys. Rev. A} \textbf{\bibinfo{volume}{91}},
  \bibinfo{pages}{042118} (\bibinfo{year}{2015}),
  \urlprefix\url{https://link.aps.org/doi/10.1103/PhysRevA.91.042118}.

\bibitem[{\citenamefont{Di~Piazza}(2016)}]{DiPiazza2016PRL}
\bibinfo{author}{\bibfnamefont{A.}~\bibnamefont{Di~Piazza}},
  \bibinfo{journal}{Phys. Rev. Lett.} \textbf{\bibinfo{volume}{117}},
  \bibinfo{pages}{213201} (\bibinfo{year}{2016}),
  \urlprefix\url{https://link.aps.org/doi/10.1103/PhysRevLett.117.213201}.

\bibitem[{\citenamefont{Jansen et~al.}(2016)\citenamefont{Jansen,
  Kami\ifmmode~\acute{n}\else \'{n}\fi{}ski, Krajewska, and
  M\"uller}}]{PRD013010}
\bibinfo{author}{\bibfnamefont{M.~J.~A.} \bibnamefont{Jansen}},
  \bibinfo{author}{\bibfnamefont{J.~Z.}
  \bibnamefont{Kami\ifmmode~\acute{n}\else \'{n}\fi{}ski}},
  \bibinfo{author}{\bibfnamefont{K.}~\bibnamefont{Krajewska}},
  \bibnamefont{and} \bibinfo{author}{\bibfnamefont{C.}~\bibnamefont{M\"uller}},
  \bibinfo{journal}{Phys. Rev. D} \textbf{\bibinfo{volume}{94}},
  \bibinfo{pages}{013010} (\bibinfo{year}{2016}),
  \urlprefix\url{https://link.aps.org/doi/10.1103/PhysRevD.94.013010}.

\bibitem[{\citenamefont{Seipt}(2017)}]{seipt2017volkov}
\bibinfo{author}{\bibfnamefont{D.}~\bibnamefont{Seipt}},
  \bibinfo{journal}{arXiv preprint arXiv:1701.03692}  (\bibinfo{year}{2017}).

\bibitem[{\citenamefont{Tang et~al.}(2020)\citenamefont{Tang, King, and
  Hu}}]{tang2020highly}
\bibinfo{author}{\bibfnamefont{S.}~\bibnamefont{Tang}},
  \bibinfo{author}{\bibfnamefont{B.}~\bibnamefont{King}}, \bibnamefont{and}
  \bibinfo{author}{\bibfnamefont{H.}~\bibnamefont{Hu}},
  \bibinfo{journal}{Physics Letters B} \textbf{\bibinfo{volume}{809}},
  \bibinfo{pages}{135701} (\bibinfo{year}{2020}), ISSN
  \bibinfo{issn}{0370-2693},
  \urlprefix\url{https://www.sciencedirect.com/science/article/pii/S0370269320305049}.

\bibitem[{\citenamefont{Tang et~al.}(2024)\citenamefont{Tang, Xin, Wen, Bake,
  and Xie}}]{MRE0196125}
\bibinfo{author}{\bibfnamefont{S.}~\bibnamefont{Tang}},
  \bibinfo{author}{\bibfnamefont{Y.}~\bibnamefont{Xin}},
  \bibinfo{author}{\bibfnamefont{M.}~\bibnamefont{Wen}},
  \bibinfo{author}{\bibfnamefont{M.~A.} \bibnamefont{Bake}}, \bibnamefont{and}
  \bibinfo{author}{\bibfnamefont{B.}~\bibnamefont{Xie}},
  \bibinfo{journal}{Matter and Radiation at Extremes}
  \textbf{\bibinfo{volume}{9}}, \bibinfo{pages}{037204} (\bibinfo{year}{2024}),
  ISSN \bibinfo{issn}{2468-2047},
  \urlprefix\url{https://doi.org/10.1063/5.0196125}.

\bibitem[{\citenamefont{Ilderton et~al.}(2020)\citenamefont{Ilderton, King, and
  Tang}}]{ILDERTON2020135410}
\bibinfo{author}{\bibfnamefont{A.}~\bibnamefont{Ilderton}},
  \bibinfo{author}{\bibfnamefont{B.}~\bibnamefont{King}}, \bibnamefont{and}
  \bibinfo{author}{\bibfnamefont{S.}~\bibnamefont{Tang}},
  \bibinfo{journal}{Physics Letters B} \textbf{\bibinfo{volume}{804}},
  \bibinfo{pages}{135410} (\bibinfo{year}{2020}), ISSN
  \bibinfo{issn}{0370-2693},
  \urlprefix\url{https://www.sciencedirect.com/science/article/pii/S0370269320302148}.

\bibitem[{\citenamefont{Jackson}(1999)}]{jackson1999classical}
\bibinfo{author}{\bibfnamefont{J.~D.} \bibnamefont{Jackson}},
  \emph{\bibinfo{title}{Classical electrodynamics}} (\bibinfo{year}{1999}).

\bibitem[{\citenamefont{Blackburn et~al.}(2020)\citenamefont{Blackburn, Seipt,
  Bulanov, and Marklund}}]{PRA2020012505}
\bibinfo{author}{\bibfnamefont{T.~G.} \bibnamefont{Blackburn}},
  \bibinfo{author}{\bibfnamefont{D.}~\bibnamefont{Seipt}},
  \bibinfo{author}{\bibfnamefont{S.~S.} \bibnamefont{Bulanov}},
  \bibnamefont{and} \bibinfo{author}{\bibfnamefont{M.}~\bibnamefont{Marklund}},
  \bibinfo{journal}{Phys. Rev. A} \textbf{\bibinfo{volume}{101}},
  \bibinfo{pages}{012505} (\bibinfo{year}{2020}),
  \urlprefix\url{https://link.aps.org/doi/10.1103/PhysRevA.101.012505}.

\bibitem[{\citenamefont{Blackburn et~al.}(2023)\citenamefont{Blackburn, King,
  and Tang}}]{PoP093903}
\bibinfo{author}{\bibfnamefont{T.~G.} \bibnamefont{Blackburn}},
  \bibinfo{author}{\bibfnamefont{B.}~\bibnamefont{King}}, \bibnamefont{and}
  \bibinfo{author}{\bibfnamefont{S.}~\bibnamefont{Tang}},
  \bibinfo{journal}{Physics of Plasmas} \textbf{\bibinfo{volume}{30}},
  \bibinfo{pages}{093903} (\bibinfo{year}{2023}), ISSN
  \bibinfo{issn}{1070-664X}, \urlprefix\url{https://doi.org/10.1063/5.0159963}.

\bibitem[{\citenamefont{Tang}(2024)}]{TANG2024139136}
\bibinfo{author}{\bibfnamefont{S.}~\bibnamefont{Tang}},
  \bibinfo{journal}{Physics Letters B} \textbf{\bibinfo{volume}{859}},
  \bibinfo{pages}{139136} (\bibinfo{year}{2024}), ISSN
  \bibinfo{issn}{0370-2693},
  \urlprefix\url{https://www.sciencedirect.com/science/article/pii/S0370269324006944}.

\bibitem[{\citenamefont{Heinzl et~al.}(2020)\citenamefont{Heinzl, King, and
  MacLeod}}]{LMA063110}
\bibinfo{author}{\bibfnamefont{T.}~\bibnamefont{Heinzl}},
  \bibinfo{author}{\bibfnamefont{B.}~\bibnamefont{King}}, \bibnamefont{and}
  \bibinfo{author}{\bibfnamefont{A.~J.} \bibnamefont{MacLeod}},
  \bibinfo{journal}{Phys. Rev. A} \textbf{\bibinfo{volume}{102}},
  \bibinfo{pages}{063110} (\bibinfo{year}{2020}),
  \urlprefix\url{https://link.aps.org/doi/10.1103/PhysRevA.102.063110}.

\bibitem[{\citenamefont{Ritus}(1985)}]{ritus85}
\bibinfo{author}{\bibfnamefont{V.~I.} \bibnamefont{Ritus}},
  \bibinfo{journal}{J. Sov. Laser Res.} \textbf{\bibinfo{volume}{6}}
  (\bibinfo{year}{1985}), \urlprefix\url{https://www.osti.gov/biblio/5972043}.

\bibitem[{\citenamefont{Gonoskov et~al.}(2015)\citenamefont{Gonoskov,
  Bastrakov, Efimenko, Ilderton, Marklund, Meyerov, Muraviev, Sergeev, Surmin,
  and Wallin}}]{2015PRE023305}
\bibinfo{author}{\bibfnamefont{A.}~\bibnamefont{Gonoskov}},
  \bibinfo{author}{\bibfnamefont{S.}~\bibnamefont{Bastrakov}},
  \bibinfo{author}{\bibfnamefont{E.}~\bibnamefont{Efimenko}},
  \bibinfo{author}{\bibfnamefont{A.}~\bibnamefont{Ilderton}},
  \bibinfo{author}{\bibfnamefont{M.}~\bibnamefont{Marklund}},
  \bibinfo{author}{\bibfnamefont{I.}~\bibnamefont{Meyerov}},
  \bibinfo{author}{\bibfnamefont{A.}~\bibnamefont{Muraviev}},
  \bibinfo{author}{\bibfnamefont{A.}~\bibnamefont{Sergeev}},
  \bibinfo{author}{\bibfnamefont{I.}~\bibnamefont{Surmin}}, \bibnamefont{and}
  \bibinfo{author}{\bibfnamefont{E.}~\bibnamefont{Wallin}},
  \bibinfo{journal}{Phys. Rev. E} \textbf{\bibinfo{volume}{92}},
  \bibinfo{pages}{023305} (\bibinfo{year}{2015}),
  \urlprefix\url{https://link.aps.org/doi/10.1103/PhysRevE.92.023305}.

\bibitem[{\citenamefont{Di~Piazza et~al.}(2018)\citenamefont{Di~Piazza,
  Tamburini, Meuren, and Keitel}}]{Piazza2018PRA012134}
\bibinfo{author}{\bibfnamefont{A.}~\bibnamefont{Di~Piazza}},
  \bibinfo{author}{\bibfnamefont{M.}~\bibnamefont{Tamburini}},
  \bibinfo{author}{\bibfnamefont{S.}~\bibnamefont{Meuren}}, \bibnamefont{and}
  \bibinfo{author}{\bibfnamefont{C.~H.} \bibnamefont{Keitel}},
  \bibinfo{journal}{Phys. Rev. A} \textbf{\bibinfo{volume}{98}},
  \bibinfo{pages}{012134} (\bibinfo{year}{2018}),
  \urlprefix\url{https://link.aps.org/doi/10.1103/PhysRevA.98.012134}.

\bibitem[{\citenamefont{Ilderton et~al.}(2019)\citenamefont{Ilderton, King, and
  Seipt}}]{king19a}
\bibinfo{author}{\bibfnamefont{A.}~\bibnamefont{Ilderton}},
  \bibinfo{author}{\bibfnamefont{B.}~\bibnamefont{King}}, \bibnamefont{and}
  \bibinfo{author}{\bibfnamefont{D.}~\bibnamefont{Seipt}},
  \bibinfo{journal}{Phys. Rev. A} \textbf{\bibinfo{volume}{99}},
  \bibinfo{pages}{042121} (\bibinfo{year}{2019}),
  \urlprefix\url{https://link.aps.org/doi/10.1103/PhysRevA.99.042121}.

\bibitem[{\citenamefont{Di~Piazza et~al.}(2019)\citenamefont{Di~Piazza,
  Tamburini, Meuren, and Keitel}}]{PiazzaPRA2019}
\bibinfo{author}{\bibfnamefont{A.}~\bibnamefont{Di~Piazza}},
  \bibinfo{author}{\bibfnamefont{M.}~\bibnamefont{Tamburini}},
  \bibinfo{author}{\bibfnamefont{S.}~\bibnamefont{Meuren}}, \bibnamefont{and}
  \bibinfo{author}{\bibfnamefont{C.~H.} \bibnamefont{Keitel}},
  \bibinfo{journal}{Phys. Rev. A} \textbf{\bibinfo{volume}{99}},
  \bibinfo{pages}{022125} (\bibinfo{year}{2019}),
  \urlprefix\url{https://link.aps.org/doi/10.1103/PhysRevA.99.022125}.

\end{thebibliography}
\providecommand{\noopsort}[1]{}

\end{document}